\documentclass[prb,showpacs,superscriptaddress,twocolumn]{revtex4}%
\usepackage{amsfonts}
\usepackage{amsmath}
\usepackage{amssymb}
\usepackage{graphicx}%
\begin{document}
\title{Ab initio study of shock compressed oxygen}
\author{Cong Wang }
\affiliation{LCP, Institute of Applied Physics and Computational
Mathematics, P.O. Box 8009, Beijing 100088, People's Republic of
China}
\author{Ping Zhang}
\thanks{Corresponding author. zhang\_ping@iapcm.ac.cn}
\affiliation{LCP, Institute of Applied Physics and Computational
Mathematics, P.O. Box 8009, Beijing 100088, People's Republic of
China} \affiliation{Center for Applied Physics and Technology,
Peking University, Beijing 100871, People's Republic of China}

\date{\today}

\begin{abstract}
Quantum molecular dynamic simulations are introduced to study the
shock compressed oxygen. The principal Hugoniot points derived from
the equation of state agree well with the available experimental
data. With the increase of pressure, molecular dissociation is
observed. Electron spin polarization determines the electronic
structure of the system under low pressure, while it is suppressed
around 30 $\sim$ 50 GPa. Particularly, nonmetal-metal transition is
taken into account, which also occurs at about 30 $\sim$ 50 GPa. In
addition, the optical properties of shock compressed oxygen are also
discussed.
\end{abstract}
\maketitle

\section{INTRODUCTION}

The study of pressure-induced transformations of materials in the
shock compressed fluid state is of great interest in the last decade
\cite{Hemley2000}. In particular, the homonuclear diatomic molecular
systems such as hydrogen \cite{Nellis1983,Knudson2004}, oxygen
\cite{Nellis1980,Hamilton1988,Bastea2001,Edwards2002}, nitrogen
\cite{Mailhiot1992,Goncharov2000}, and the halogens
\cite{Balchan1961,Reichlin1994}, have been intensively studied with
a variety of experimental techniques, and the measurements of
Hugoniots have reached megabar range. On the theoretical side,
Quantum molecular dynamic (QMD) and path-integral Monte Carlo (PIMC)
simulations \cite{Militzer2003,Kress2000,Militzer2009} have been
introduced in the study of materials under extreme conditions. In
spite of fact that these theoretical simulations in some cases
already provided explanatory and predictive results, to date,
however, many fundamental questions of these above-mentioned systems
in the fluid state are still under intense discussion due to the
profound behavior they display.

A knowledge of the properties of oxygen, in particular, the equation
of state (EOS) and derived properties, such as the Hugoniot curve
and ($D,u$) diagram, are important for shocks, detonations, biology,
and fluids. Single or multiple shock-wave experiments have been
performed on oxygen \cite{Nellis1980,Hamilton1988}, and principal
Hugoniot are determined from the EOS data. The partial molecular
dissociation and a two-component conductive fluid are indicated by
the results. Theoretically, a classical repulsive pair potential was
introduced to investigate the EOS of shock-compressed oxygen
\cite{Ross1980}, and the dissociation effect of oxygen at high
pressure was partially taken into account by using a classical
self-consistent fluid variational theory \cite{Chen2008}. Due to the
intrinsic approximations of these classical methods, however, a full
quantum-mechanical description of the change in the electronic
structure of oxygen under high-pressure and high-temperature
shock-wave compression still remains to be presented and understood.
The necessity for such quantum-mechanical treatment of shock
compressed oxygen can, for example, be clearly seen by the following
facts: (i) QMD method, where the electrons are fully quantum
mechanical treated, has been proven successfully in calculating the
electronic structure, thermodynamical and optical properties of warm
dense matter \cite{Mattsson2006,Kietzmann2007}; (ii) nonmetal-metal
transition of oxygen was observed recently \cite{Bastea2001}; (iii)
low-pressure QMD calculations \cite{Militzer2003} already indicated
importance of electron spin polarization effect during oxygen
dissociation. From these facts, therefore, electronic structure
should be seriously considered, the EOS data and Hugoniot points
from density functional theory (DFT) are highly needed for shock
compressed oxygen.

In the present work, we apply QMD simulations to study the
thermophysical properties of shock compressed oxygen. We determine
the EOS data and dissociation fraction of oxygen along the principal
Hugoniot. The pair correlation function is derived from the
structural information during QMD simulations. The electronic
structure calculations within QMD provide the charge density in the
simulation box and density of states (DOS) at every time step. The
dissociation of oxygen and electronic structure are important for
identifying nonmetal-metal transition. The dielectric function
$\varepsilon(\omega)$ and the consequent optical quantities are also
extracted.

\section{QUANTUM MOLECULAR DYNAMICS SIMULATIONS}

In this study, we employ the VASP plane-wave pseudopotential code,
which was developed at the Technical University of
Vienna\cite{Kresse1993,Kresse1996}. The finite temperature density
functional theory molecular dynamics (FTDFT-MD)
\cite{Lenosky2000,Bagnier2001}, where the electronic states are
populated according to a Fermi-Dirac distribution at temperature
T$_{e}$, is used in the present work. The electron wave functions
are calculated with the projector augmented wave (PAW) potential
\cite{Blochl1994,Kresse1999}. The exchange correlation functional is
calculated within generalized gradient approximation (GGA) using the
parametrization of Perdew-Wang 91 \cite{Perdew1991}. Electron spin
polarization, which can be effectively used to describe the
character of O$_{2}$ molecules, is also considered in our
calculation.

In all the simulations, a total number of 96 oxygen atoms (48
O$_{2}$ molecules) are included in a supercell with periodic
boundary condition. Trajectories are calculated at separate
densities and temperatures. The selected densities range from
$\rho$=2.5 to 3.6 g/cm$^{3}$ with temperatures from 1000 to 16000 K.
The plane wave cutoff energy is set to be 600.0 eV.  For molecular
dynamic simulations, only $\Gamma$ point of the Brillouin zone is
included, while $4\times4\times4$ Monkhorst-Pack
\cite{Monkhorst1976} $k$ points are used for calculating the
electronic structure. Integration of equations of motion proceeds
2000 steps with the time step of 2 fs. Then, the system is
equilibrated 200 steps and the final 300 steps are used to calculate
physical quantities. During simulations, the ionic temperature
$T_{i}$ is kept constant at every time step by using velocity
scaling. The local thermodynamical equilibrium is reached by setting
electron temperature $T_{e}$ equal to $T_{i}$. The accuracy of our
calculations is examined by the bond length of O$_{2}$ molecule in
its ground state, and the result is 1.24 \AA, which agrees well with
the experiment \cite{Huber1979}.

\section{PRINCIPAL HUGONIOT AND PAIR CORRELATION FUNCTION}

A crucial measurement of the EOS data of oxygen under shock
conditions is the Hugoniot \cite{Zeldovich1966} which is the locus
of points in ($E, P, V$)-space satisfying the condition:
\begin{equation} \label{hugoniot}
    (E_{0}-E_{1})+\frac{1}{2}(V_{0}-V_{1})(P_{0}+P_{1})=0,
\end{equation}
where $E$ is the internal energy, $P$ is the pressure, $V$ is the
volume, and the subscripts 0 and 1 denote the initial and shocked
state, respectively. This relation follows from conservation of
matter, momentum, and energy for an isolated system compressed by a
pusher at a constant velocity. In the canonical (NVT) ensemble in
which both $E$ and $P$ are temperature dependent, the locus of
states which satisfy Eq. (1) is the so-called principal Hugoniot,
which describes the shock adiabat between the initial and final
states. In our first-principles calculations, the internal energy
consists of the DFT total energy and the zero-point vibrational
energy ($\frac{1}{2}h\nu_{vib}$ per molecule). The pressure is
evaluated via the forces provided by VASP. The particle velocity
$u_{p}$ and shock velocity $u_{s}$ can be derived from the other two
Rankine-Hugoniot equations \cite{Zeldovich1966}
$V_{1}=V_{0}[1-(u_{p}/u_{s})]$ and $P_{1}-P_{0}=\rho_{0}u_{s}u_{p}$.
In order to calculate the principal Hugoniot point for a given
$V_{1}$, a series of simulations are executed for different
temperatures $T$. ($E_{0}-E_{1}$) and
$\frac{1}{2}(V_{0}-V_{1})(P_{0}+P_{1})$ are then fitted to
polynomial expansion of $T$. The principal Hugoniot temperature
$T_{1}$ and pressure $P_{1}$ are then determined by solving Eq. (1).

\begin{figure}[ptb]
\begin{center}
\includegraphics[width=1.0\linewidth]{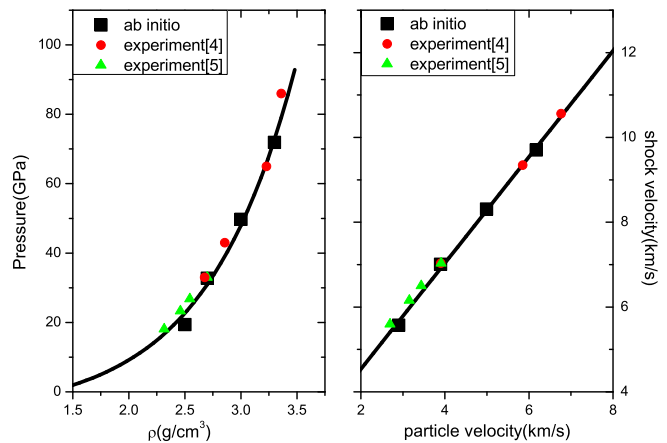}
\end{center}
\caption{(Color online). Left panel: Calculated principal Hugoniot
curve, in comparison with the previous experimental data; Right
panel: $u_{p}$-$u_{s}$ curve.}%
\label{fig1}%
\end{figure}

\begin{table}[!htbp]
\centering \caption{Principal Hugoniot points derived from DFT-MD
simulation at a series of density ($\rho$), pressure ($P$), and
temperature ($T$). The corresponding particle velocity ($u_{p}$),
shock velocity ($u_{s}$) and dissociation ratio ($R$) are included.}
\begin{tabular}{cccccc}
\hline\hline
$\rho$ (g/cm$^{3}$)& $P$ (GPa) & $T$ (K) & $u_{p}$ (km/s) & $u_{s}$ (km/s) & $R$\\
\hline
 2.5  & 19.4  &  2413 & 2.90 & 5.57 & 27\% \\
 2.7  & 32.8  &  5258 & 3.89 & 7.01 & 54\% \\
 3.0  & 49.8  &  6489 & 4.99 & 8.31 & 69\% \\
 3.3  & 71.9  &  8823 & 6.18 & 9.70 & 71\% \\
\hline\hline
\end{tabular}
\label{H_data}
\end{table}

\begin{figure}[ptb]
\begin{center}
\includegraphics[width=1.0\linewidth]{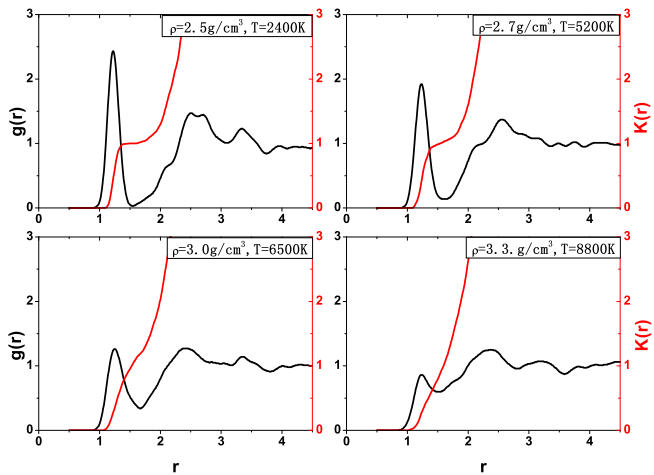}
\end{center}
\caption{(Color online). Pair correlation functions (black line) and
$K(r)$ (red line) along the principal Hugoniot curve.}%
\label{fig2}%
\end{figure}

As the starting point of the principal Hugoniot curve, it is
important to determine the initial state ($E_{0}$, $P_{0}$,
$V_{0}$). Generally, the initial-state internal energy should be
calculated under the experimental condition, with $\rho$=1.2
g/cm$^{3}$ and $T$=77 K. However, this calculation is highly time
demanding and is out of our computational capabilities. As an
alternative selection, we determine the initial-state energy via
single molecule energy. In the present work, we perform PAW
potential calculation of an isolated (spin-polarized) O$_{2}$
molecule at the equilibrium bond length. With the inclusion of zero
point energy, the internal energy $E_{0}$ is $-$4.84 eV/atom at
$T$=77 K. The initial pressure $P_{0}$ can be neglected compared
with the high pressures of shock compressed states.

The principal Hugoniot curve and ($u_{s}$, $u_{p}$) diagram are
shown in Fig. 1. We find good agreement between our DFT-MD results
and the previous experimental data \cite{Nellis1980,Hamilton1988}.
The present DFT-MD simulations can reasonably reflect the main
tendency of oxygen thermodynamic properties along the principal
Hugoniot curve. We now investigate explicit nature of the fluid for
which the experiments provide only indirect evidence. The
probability of finding a atom at distance $r$ from a reference atom
could be given by pair correlation function $g(r)$, which is
obtained by averaging over all particles and simulation steps in
equilibrium. The fraction of molecule can be derived by evaluating
the coordination number, which could be expressed as follows:
\begin{equation} \label{fraction}
K(r)=\frac{N-1}{V}\int_{0}^{r}4\pi r'^{2}g(r')dr'.
\end{equation}
The coordination number is a weighted integral over the pair
correlation function $g(r)$ of the ions. $N$ is the total number of
ions and $V$ is the volume of the supercell. The value $K$ at the
maximum of $g(r)$ (around 1.24 \AA) is equal to the fraction of
molecules in the supercell. The calculated Hugoniot points and
related data, such as temperatures and dissociation fraction, are
listed in Table I.

Figure 2 shows the pair correlation functions and $K(r)$ for four
densities and temperatures. At the density $\rho$=2.5 g/cm$^{3}$ and
T=2400 K, the pair correlation function is featured by a sharp
maximum peak around the equilibrium bond length of 1.24
 \AA, and $K(1.24)=73\%$. Under this condition, the pair
correlation function exhibits the typical characteristics of a
molecular fluid. As pressure increases, the first main peak begin to
reduce in amplitude. This tendency means that the dissociation of
O$_{2}$ molecules becomes remarkable. The molecule fraction is 46\%
at $\rho$=2.7 g/cm$^{3}$ and 31\% at $\rho$=3.0 g/cm$^{3}$. The
crossover from molecular to atomic fluid exists around $\rho$=2.7
$\sim$ 3.0 g/cm$^{3}$, and the corresponding principal Hugoniot
pressure range from 30 to 50 GPa. Under such conditions, O$_{2}$
molecules dissociate gradually.

\section{ELECTRONIC STRUCTURE}

Metallization of oxygen was first reported by Desgreniers $et$ $al.$
\cite{Desgreniers1990} in solid state, with the nonmetal-metal
transition accomplished by structural change \cite{Weck2002}. The
nonmetal-metal transition of oxygen in fluid state was later
reported by Bestea $et$ $al.$ \cite{Bastea2001}. The ground state of
oxygen is a spin-triplet state under low pressures. The molecular
structure and magnetism properties mainly originate from electron
spin polarization \cite{Oda2007}. Therefore, the electronic
structure, which dominates the physical properties, is examined
along the principal Hugoniot curve.

\begin{figure}[ptb]
\begin{center}
\includegraphics[width=0.6\linewidth]{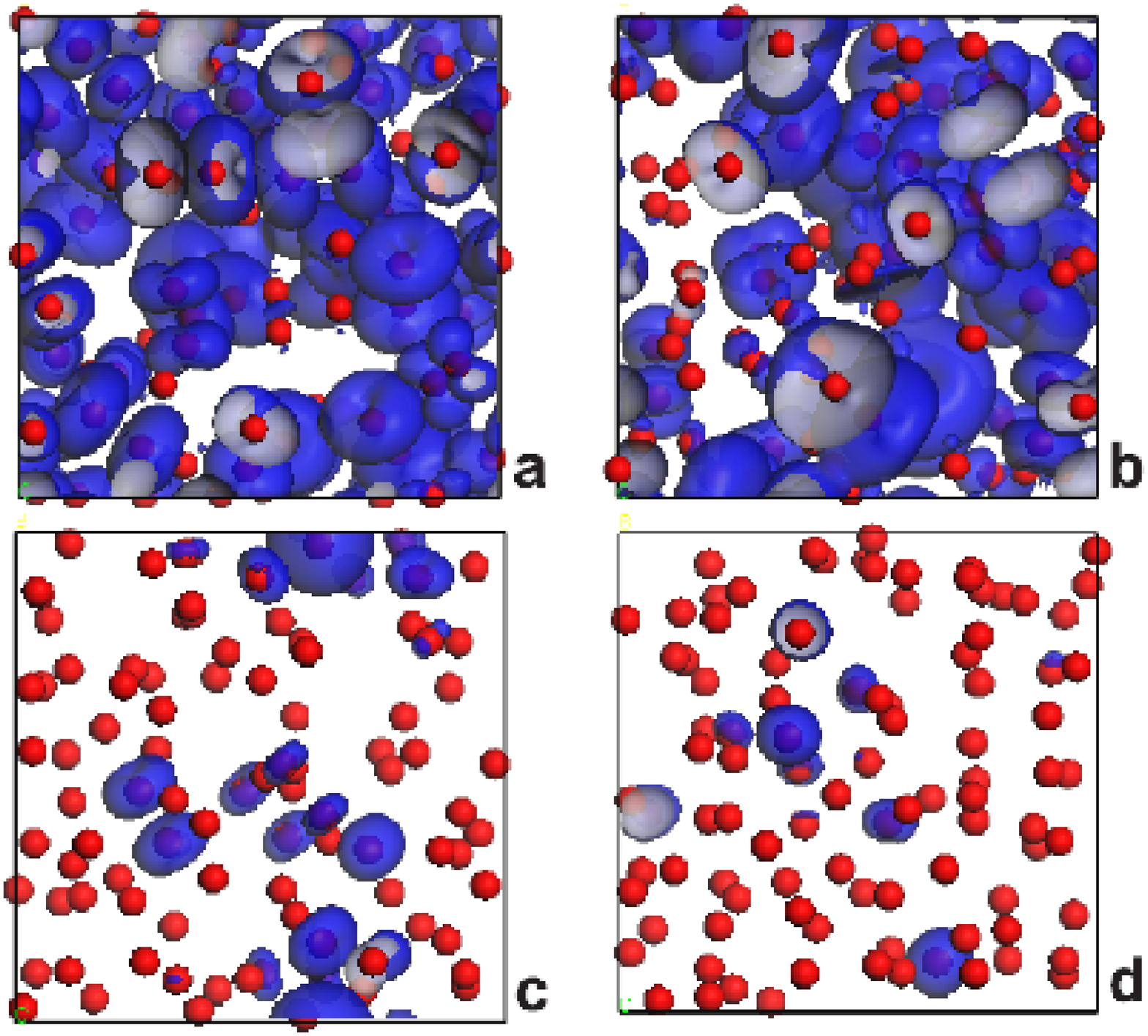}
\end{center}
\caption{(Color online). Isosurface of spin density for (a)
$\rho$=2.5 g/cm$^{3}$, $T$=2400 K, (b) $\rho$=2.7 g/cm$^{3}$,
$T$=5200 K, (c) $\rho$=3.0 g/cm$^{3}$, T=6500 K, and (d) $\rho$=3.3 g/cm$^{3}$, $T$=8800 K.}%
\label{fig3}%
\end{figure}

Figure 3 shows the isosurfaces of spin densities of oxygen at the
four principal Hugoniot points. At low pressures ($P=20\sim$ 30
GPa), the fluid mainly consists of disordered O$_{2}$ molecules, and
less than 50\% fraction of molecules dissociate. In this case, the
finite spin polarization characteristic of molecular oxygen is
obvious, as shown in Fig. 3(a). As a result, the structural,
electrical, and optical properties of the system is closely related
with the electron spin polarization of oxygen molecules at low
pressures. With the increase of pressures ($P=50\sim$ 70 GPa),
electron spin polarization is suppressed, and the spin density in
the supercell tends to disappear, as shown in Figs. 3(c) and 3(d).
This change will consequently affect DOS of the system, which plays
a key role in elucidating nonmetal-metal transition and optical
properties. As shown in Fig. 4, which plots spin-resolved DOS of
shocked oxygen in different pressure regions, a large band gap
exists in the DOS when the pressure is about 20GPa ($\rho$=2.5
g/cm$^{3}$, $T$=2400 K), and the system stays in the insulating
state. With the increase of pressure, the electrons tend to be
delocalized by the non-zero value of DOS at the Fermi energy.
Finally, the localized molecular bonds as well as the
valence-conduction band gap prominently disappear and the metallic
like conductivity emerges at this stage. The nonmetal-metal
transition, which is attributed to the dissociation of molecules and
thermal broadening of the electronic states, occurs at the pressure
range from 30 to 50 GPa. This result agrees well with the experiment
\cite{Hamilton1988}.

\begin{figure}[ptb]
\begin{center}
\includegraphics[width=1.0\linewidth]{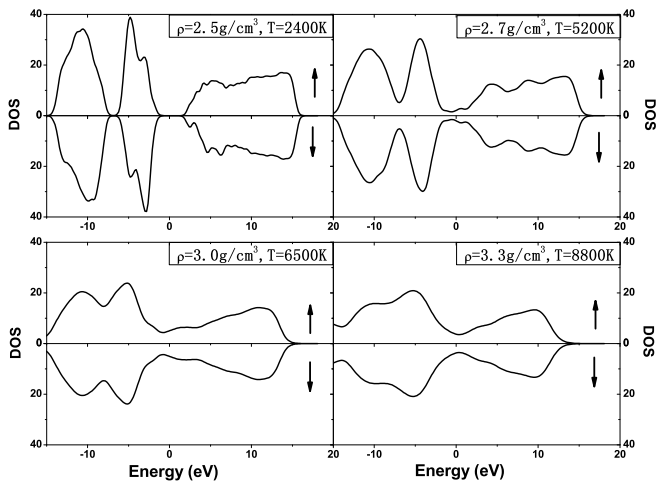}
\end{center}
\caption{Spin-resolved electron densities of states along the
principal Hugoniot curve. The Fermi energy is set to be zero. (a)
$\rho$=2.5g/cm$^{3}$, $T$=2400 K; (b) $\rho$=2.7 g/cm$^{3}$,
$T$=5200 K; (c) $\rho$=3.0
g/cm$^{3}$, $T$=6500 K; (d) $\rho$=3.3 g/cm$^{3}$, $T$=8800 K.}%
\label{fig4}%
\end{figure}

\section{OPTICAL PROPERTIES}

Having shown the EOS data and nonmetal-metal transition along the
principal Hugoniot curve, we turn now to study the optical
properties of warm dense oxygen. The frequency dependent dielectric
function $\epsilon(\omega)$ has both real and imaginary parts:
\begin{equation} \label{dielectric}
\epsilon(\omega)=\epsilon^{(1)}(\omega)+\epsilon^{(2)}(\omega)i.
\end{equation}
The real $n(\omega)$ and imaginary $k(\omega)$ parts of the
refraction index are related to the dielectric function by
\begin{equation} \label{refraction}
\epsilon(\omega)=\epsilon^{(1)}(\omega)+\epsilon^{(2)}(\omega)i=[n(\omega)+k(\omega)i]^{2},
\end{equation}
or equivalenly
\begin{equation} \label{refraction-real}
n(\omega)=\sqrt{\frac{1}{2}[|\epsilon(\omega)|+\epsilon^{(1)}(\omega)]},
\end{equation}

\begin{equation} \label{refraction-imaginary}
k(\omega)=\sqrt{\frac{1}{2}[|\epsilon(\omega)|-\epsilon^{(1)}(\omega)]}.
\end{equation}
From the refraction index, the reflectivity $r(\omega)$ is given by
\begin{equation} \label{reflectivity}
r(\omega)=\frac{[1-n(\omega)]^{2}+k(\omega)^{2}}{[1+n(\omega)]^{2}+k(\omega)^{2}}.
\end{equation}

In Fig. 5 (left panel), we show the variation of the reflectivity
$r(\omega)$ along the principal Hugoniot curve. At lower pressures,
for example P=20$\sim$30 GPa, there are two peaks around the photon
energies of 5.0 and 11.0 eV, corresponding to the wavelength of 250
and 110 nm, respectively. With the increase in pressure, the peaks
vanish. The reflectivity reaches a constant value of 0.172 for the
photon energy higher than 35.0eV. We select a typical wavelength of
414 nm to investigate the change in reflectivity, as shown in the
right panel of Fig. 5. One can see that the reflectivity increases
from 0.01 to 0.2 along the principal Hugoniot. The rapid increase in
reflectivity occurs at the pressure range from 30 to 50 GPa, with
about 50\% dissociation of molecules.

\begin{figure}[ptb]
\begin{center}
\includegraphics[width=1.0\linewidth]{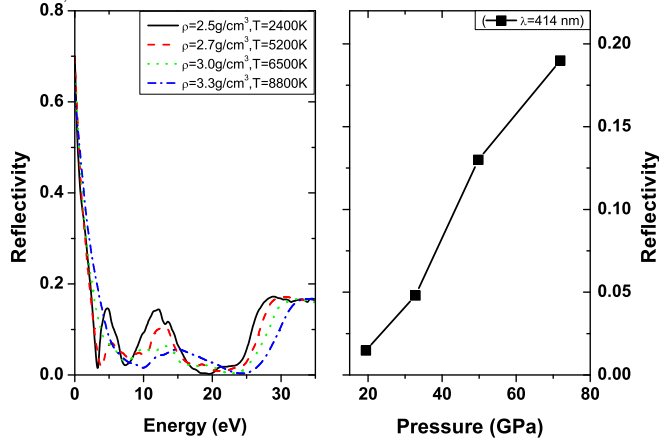}
\end{center}
\caption{(Color online) Variation of the frequency-dependent optical
reflectivity along the principal Hugoniots (left panel);
Reflectivity at a fixed
wavelength of 414 nm (right panel).}%
\label{fig5}%
\end{figure}

\section{CONCLUSION}

In summary, we have applied QMD simulations in the study of
shock-compressed oxygen. The principal Hugoniot curve has been
derived from the computational EOS data, which is consistent well
with the experimental measurement. The electron spin polarization,
determines physical properties of oxygen at lower pressures, has
been shown to be greatly suppressed at higher pressures. This fact,
as well as the nonmetal-metal transition and change in reflectivity
found in this paper, are attributed to the dissociation of O$_{2}$
molecules. We have determined that the molecular-atomic fluid
transition exists at the pressure around 30$\sim$50 GPa, which also
accord well with the recent experiment.

\begin{acknowledgments}
This work was supported by the Foundation for Development of Science
and Technology of China Academy of Engineering Physics under Grant
No. 2009B0301037.
\end{acknowledgments}

\end{document}